# Channel Knowledge Maps for 6G Wireless Networks: Construction, Applications, and Future Challenges

Xingchen Liu[1], Shu Sun[1], Meixia Tao[1], Aryan Kaushik[2], Hangsong Yan[3]
1. School of Information Science and Electronic Engineering, Shanghai Jiao Tong University, Shanghai 200240, China;
2. Department of Computing and Mathematics, Manchester Metropolitan University, Manchester M15 6BH, UK;
3. Hangzhou Institute of Technology, Xidian University, Hangzhou 311231, China

Corresponding author: Shu Sun (email: shusun@sjtu.edu.cn)

**Abstract:**
The advent of 6G wireless networks promises unprecedented connectivity, supporting ultra-high data rates, low latency, and massive device connectivity. However, these ambitious goals introduce significant challenges, particularly in channel estimation due to complex and dynamic propagation environments. This paper explores the concept of channel knowledge maps (CKMs) as a solution to these challenges. CKMs enable environment-aware communications by providing location-specific channel information, reducing reliance on real-time pilot measurements. We categorize CKM construction techniques into measurement-based, model-based, and hybrid methods, and examine their key applications in integrated sensing and communication systems, beamforming, trajectory optimization of unmanned aerial vehicles, base station placement, and resource allocation. Furthermore, we discuss open challenges and propose future research directions to enhance the robustness, accuracy, and scalability of CKM-based systems in the evolving 6G landscape.

## 1 Introduction

The advent of 6G wireless networks promises to revolutionize connectivity by enabling ultra-high data rates, ultra-low latency, massive device connectivity, and pervasive intelligence[1–3]. However, these ambitious performance targets come with significant challenges. The increasing density of devices and network nodes[4], along with the deployment of large-scale antenna arrays[5] and utilization of higher frequency bands[6] (such as millimeter-wave (mmWave) and terahertz (THz)), leads to highly complex propagation environments. In such scenarios, traditional channel estimation methods, primarily based on extensive pilot training, are rapidly becoming inefficient, as they struggle to cope with the increased channel dimensions and dynamic variability in the environment[7]. At the same time, the wealth of diverse environmental data and the rapid evolution of artificial intelligence techniques offer unprecedented opportunities to rethink how wireless channels are characterized and managed[8].

To address these challenges, researchers have proposed novel approaches that leverage both environmental information and advanced data analytics to enhance channel state information acquisition. One promising paradigm, first introduced in Ref. [9], is the channel knowledge map (CKM), which is a site-specific database that links geographical locations to detailed channel parameters. By exploiting the spatial consistency inherent in wireless propagation, CKMs enable networks to infer channel characteristics based on location data, thereby reducing the dependency on real-time, high-overhead pilot measurements[10–11]. This innovative concept not only promises to alleviate the challenges posed by dense and dynamic 6G environments but also paves the way for proactive and predictive communication strategies that can significantly enhance network performance.

A CKM can be classified into different types based on the scope and granularity of channel knowledge it provides. Some CKMs focus on large-scale channel characteristics, such as path loss and shadowing, which are primarily influenced by terrain and urban structures. Others capture small-scale fading properties, including multipath effects characterized by parameters like the angle of arrival (AoA), the angle of departure (AoD), time delay, and the Doppler shift. Additionally, CKMs can be categorized based on their coverage range, such as base station-to-any (B2X) CKM, which maps the channel characteristics from a given base station to any receiver within a coverage area, or any-to-any (X2X) CKM, which generalizes channel relationships among arbitrary locations.

The construction of CKMs generally falls into three broad approaches: measurement-based methods, model-based methods, and hybrid data-model approaches. Measurement-based methods rely on empirical measurements and employ spatial interpolation techniques including the nearest-neighbor interpolation[12] and geostatistical methods like Krigin[13]. In contrast, model-based methods use well-established propagation models, including empirical formulas (e.g., COST-231 Hata[14]) and deterministic ray tracing[15]. To address the limitations of both approaches, hybrid methods combine model predictions with real-world measurements, often leveraging advanced machine learning techniques such as deep neural networks and generative adversarial networks (GANs) to enhance CKM's accuracy and adaptability.

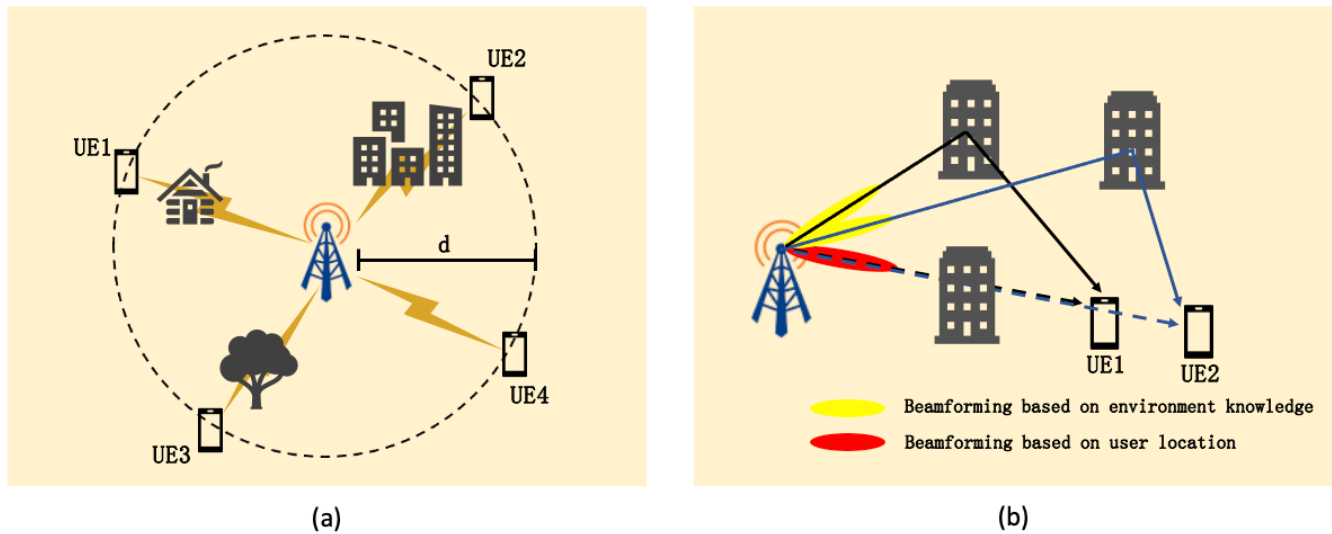

Figure 1. Illustration of environment knowledge enabled by channel knowledge map[9]: (a) path loss prediction and (b) beamforming

Based on the provided image, the key idea behind CKM is not just about specific applications but rather a paradigm shift in wireless communication towards environment-aware communications. Fig. 1 illustrates this transformation based on two examples from Ref. [9], by comparing conventional location-based or probabilistic channel modeling with environment knowledge. In Fig. 1a, CKM enables improved path loss prediction by considering environmental obstructions, rather than relying solely on distance-based models. In Fig. 1b, it enhances beamforming by incorporating environment knowledge, allowing for more accurate signal directionality and avoiding obstacles that would otherwise degrade communication performance.

This article is structured as follows. Section 2 introduces the fundamental concepts of CKMs, including their definitions, core principles, and roles in enabling environment-aware communications in 6G networks. Section 3 discusses CKM construction techniques, categorized into measurement-based, model-based, and hybrid data-model approaches, highlighting their methodologies and trade-offs. Section 4 explores key applications of CKMs, such as localization and sensing systems, trajectory optimization, beamforming and base station placement. Section 5 discusses open challenges and outlines future research directions and the transformative potential of CKMs in next-generation wireless systems. Section 6 concludes this article with a summary of the current state of CKM research, emphasizing its potential to revolutionize wireless communications by enabling proactive, environment-aware systems and addressing the challenges that lie ahead in the rapidly evolving 6G landscape.

## 2 Fundamental Concepts of Channel Knowledge Map

A CKM is a channel knowledge database associated with specific geographical locations, which is constructed to provide region-specific or location-specific channel information, thereby enhancing the understanding of the wireless propagation environment. Mathematically, a CKM can be defined as a function that maps a location vector $q \in R^D$ to a channel knowledge vector $z \in C^J$, where $q$ represents the location of the transmitter and/or receiver, and $z$ the relevant channel knowledge.

$$M: R^D \to C^J \qquad (1).$$

This knowledge may include, but is not limited to, path gain, multipath propagation parameters (such as angles of arrival, angles of departure, delay, and Doppler shifts), and the complete channel impulse response.

Intuitively, the environmental awareness capability of CKM is derived from the fundamental observation that when a device revisits a previously accessed location, it experiences a wireless propagation environment that is highly similar to the past. By fully utilizing the trajectory information of devices and surrounding environmental data, CKM can significantly reduce channel uncertainty, thereby enabling more accurate channel inference and effective communication strategies.

**2.1 Channel Modeling in a Given Region**
In a specific geographical area, the wireless propagation channel $z(t)$ is fundamentally a function of the device's position $q(t)$ and the surrounding wireless environment $E(t)$:
$$z(t) = f(q(t), E(t)) \qquad (2),$$
where $E(t)$ represents the propagation environment, which consists of both static and dynamic components. The static environment includes terrain, building structures, and material properties, while the dynamic environment accounts for moving objects such as vehicles and pedestrians. However, due to the complex interactions between electromagnetic waves and the surrounding environment, deriving the function $f(\cdot, \cdot)$ analytically is extremely difficult. Additionally, representing the environment $E(t)$ in a mathematically tractable form is non-trivial due to its high-dimensional and dynamic nature.

To overcome these challenges, CKM leverages historical data to model channel knowledge without requiring an explicit expression of the function $f(\cdot, \cdot)$ or the environmental representation $E(t)$. In a quasi-static environment, where $E(t) \approx E$, the channel can be rewritten as:
$$z(t) = f(q(t), E) \qquad (3).$$
By collecting a set of historical location data $\{q_1, q_2, \ldots, q_Q\}$ and their corresponding channel knowledge $\{z_1, z_2, \ldots, z_Q\}$, CKM can model the environment as:
$$E = g(q_{i=1,\ldots,Q}, z_{i=1,\ldots,Q}) \qquad (4),$$
which enables the inference of location-specific channel knowledge based on prior measurements.

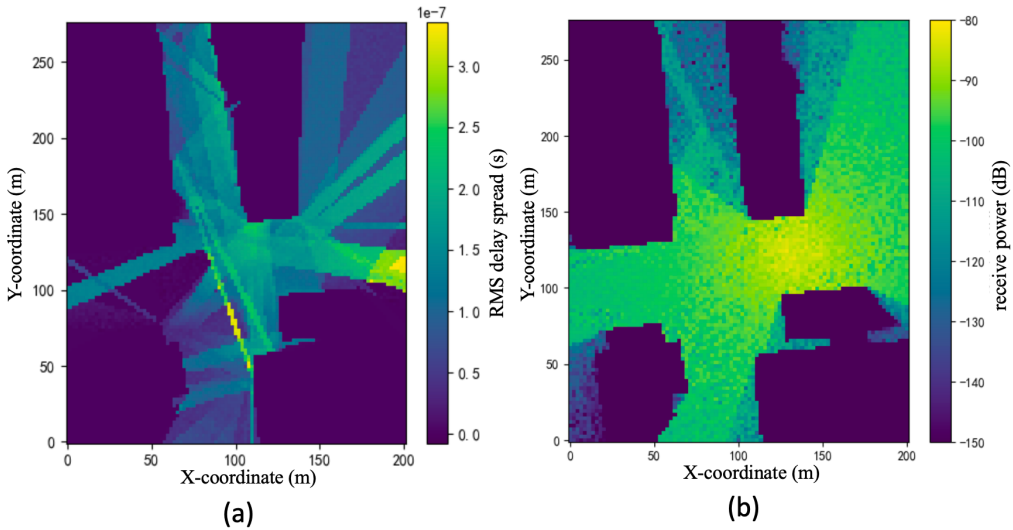

RMS: root mean square
Figure 2. Example of channel knowledge map representation in a given region:
(a) RMS delay spread map and (b) received power map

Fig. 2 illustrates CKM construction for a specific geographic area, using the calculation result of Wireless Insite, a widely used ray tracing simulation software. Fig. 2a shows the root mean square (RMS) delay spread, which reflects how obstacles, buildings for example, influence multipath propagation. In areas with dense buildings or structures, the signal experiences multiple reflections and scattering, leading to higher delay spread. Conversely, open areas with fewer obstructions result in lower delay spread, indicating less interference and multipath. These variations in delay spread are primarily due to the distribution and density of obstacles, which affects how signals propagate across the area. Fig. 2b presents the received power distribution, demonstrating how environmental factors contribute to signal attenuation. Regions with dense buildings show higher attenuation due to diffraction and reflection, leading to lower received power. On the other hand, open areas with fewer obstacles allow the signal to propagate more freely, resulting in higher received power. Together, these visualizations highlight how the distribution and arrangement of obstacles significantly affect both the delay spread and received power, illustrating the spatial consistency of wireless channels.

**2.2 Advantages and Challenges of CKM over Traditional Communication Methods**
Compared with conventional environment-agnostic communications, CKM offers several distinct advantages by providing location-aware channel knowledge. Firstly, by leveraging pre-stored historical data, CKM significantly reduces the reliance on real-time channel estimation, minimizing training overhead and improving spectral efficiency. Also, by integrating both location and environmental information, CKM improves the accuracy of channel knowledge inference, particularly in complex propagation environments and ultra-dense network deployments. Furthermore, CKM enhances network robustness and adaptability by enabling proactive communication strategies. Traditional wireless systems reacting to real-time channel variations often leads to inefficiencies in mobility management and resource allocation. In contrast, CKM allows networks to anticipate channel conditions based on environmental awareness, facilitating preemptive beam adjustments, intelligent handovers, and optimized power control.

However, it is important to note that the effectiveness of CKM in mobile scenarios depends on its ability to continuously update and adapt to real-time conditions. Although CKM reduces the need for frequent real-time pilot measurements, it still requires periodic updates to maintain its accuracy, especially in highly dynamic environments with rapid device movement. This introduces the challenge of balancing the trade-off between maintaining real-time accuracy and reducing the overhead associated with constant updates. In some cases, the accuracy of CKM may be lower than that of traditional real-time channel estimation, particularly in rapidly changing environments where instantaneous channel state information is crucial. Therefore, while CKM improves efficiency by reducing the reliance on frequent channel estimations, its accuracy may not always surpass that of traditional systems, especially when the propagation environment undergoes significant and rapid changes.

Nevertheless, CKM can outperform traditional real-time channel estimation in scenarios where efficiency and resource optimization are critical, such as in large-scale networks or in environments where real-time measurements are expensive or impractical. Moreover, integrating CKM with real-time data through adaptive learning mechanisms or sensor fusion techniques could enhance its accuracy, allowing it to approach the performance of real-time estimation in dynamic environments.

**2.3 Role of CKM in 6G Environment-Aware Communications**

In summary, CKM leverages spatial consistency and historical channel knowledge to provide an efficient and scalable solution to 6G wireless communications. Transitioning from environment-agnostic communication to environment-aware communication, CKM represents a paradigm shift in how channel knowledge is acquired and utilized. This novel approach lays the foundation for proactive and predictive communication strategies, ultimately improving the efficiency and robustness of next-generation wireless systems.

# 3 CKM Construction

**3.1 Measurement-Based CKM Construction**

Measurement-based methods for CKM construction rely entirely on empirical measurements to estimate channel characteristics across a given region. These methods utilize interpolation and regression techniques to infer the channel parameters at unmeasured locations, assuming spatial correlation in the wireless propagation environment. By avoiding explicit propagation models, these approaches capture real-world channel variations more effectively. For instance, Kriging interpolation has been widely used to create channel maps by incorporating spatial correlations in measurement data[13]. Similarly, *k*-nearest neighbor (KNN) methods have been applied to estimate channel conditions at unmeasured points based on the nearest available measurements[16].

Relying on real-world data rather than theoretical models, measurement-based CKMs can more accurately reflect actual conditions in a given environment. However, the success of these systems depends on the density and quality of measurement data, as well as the ability to effectively interpolate or extrapolate that data to areas not directly measured. Some studies have demonstrated the effectiveness of these methods in urban and rural environments, but challenges remain in areas with sparse or highly variable data[13].

1) KNN interpolation
KNN interpolation is a simple yet effective approach to estimating unknown channel values in a CKM. Given a target location $q_0$, the estimated channel knowledge $f(q_0)$ is computed as a weighted average of the $k$-nearest known measurements $z_k$: $f(q_0) = \sum k \in \mathcal{N}(q_0) w_k z_k$, where $\mathcal{N}(q_0)$ is the set of $k$ nearest measurements based on the smallest Euclidean distance $|q_0 - q_i|$, and $w_k$ is the weight assigned to each measurement. The weight can be determined using the inverse distance weighting (IDW) rule, where closer measurements contribute more significantly to the estimate. Alternatively, a kernel function can be used to define $w_k$, such as the Gaussian kernel: $w(\boldsymbol{q_0}, \boldsymbol{q_k}) = c \cdot \exp(-|\boldsymbol{q_0} - \boldsymbol{q_k}|/\sigma)$, or the Laplacian kernel: $w(q_0, q_k) = c \cdot \exp(-|q_0 - q_k|/\sigma)$. These kernel functions ensure a smooth interpolation by emphasizing the influence of nearby measurements while reducing the contribution of distant points.

2) Kriging interpolation
Kriging is a geostatistical interpolation method that estimates unknown channel values based on spatially correlated measurements[13, 17], providing an optimal linear unbiased prediction. Unlike simple interpolation techniques, Kriging leverages the spatial structure of the data through the semivariogram, which describes the degree of correlation between two points as a function of their separation distance. Given a set of known measurement locations $\{q_i, z_i\}$, Kriging estimates the channel value at an unknown location $q$ as: $\hat{f}(q) = \sum_{i=1}^{N} \lambda_i z_i + \lambda_0$, where $\lambda_i$ is the interpolation weights chosen to minimize the mean squared error (MSE), and $\lambda_0$ is a bias term to account for the mean of the underlying random process. The weights are computed by solving a linear system derived from the semivariogram function: $\gamma(q_i, q_j) = \frac{1}{2} E\left[(z_i - z_j)^2\right]$, which captures how the variance of channel measurements evolves with distances. This enables Kriging to make statistically sound predictions with quantified uncertainty.

Kriging is widely applied in CKM construction to develop interference maps[18], which capture spatial variations in interference power. By interpolating interference measurements from multiple devices, these maps support interference-aware resource allocation, allowing networks to optimize transmission parameters and minimize co-channel interference. Similarly, shadowing maps generated via Kriging[19] model large-scale signal fluctuations caused by environmental obstructions. In cognitive radio networks, they help secondary users estimate interference from primary users for efficient spectrum access. In heterogeneous networks, they aid coverage prediction and power control, improving communication reliability. Through these applications, Kriging enhances environment-aware CKM construction, enabling more adaptive and intelligent wireless systems.

3) Matrix completion
Matrix and tensor completion techniques are powerful tools for constructing CKMs when channel measurements are sparse or incomplete[20–22]. These methods leverage the low-rank structure of wireless channel data to infer missing values, reducing the need for extensive measurements while maintaining accuracy. In CKM construction, the channel knowledge across a region can be represented as a matrix $\mathbf{Z} \in R^{M \times N}$, where missing entries are estimated using matrix completion methods like nuclear norm minimization and alternating least squares. When extended to multi-dimensional data, tensor completion methods (for example, tensor nuclear norm minimization) incorporate additional factors such as frequency and time, enhancing CKM's predictive capabilities. These techniques enable efficient CKM updates, dynamic channel estimation, and resource allocation, making them valuable for real-world wireless systems with limited measurement availability.

4) Other methods
Radial basis function (RBF) interpolation[23] is a widely applied technique that estimates unknown channel values by fitting a smooth function to known measurements, ensuring spatial continuity in CKM. Gaussian process regression models the channel as a Gaussian process with a spatial covariance function[24–25], providing both predictions and uncertainty quantification, which makes it particularly useful for adaptive measurement strategies. Thin plate splines interpolation is another effective approach that minimizes bending energy to produce smooth surface reconstructions[26], capturing gradual variations in channel characteristics. These methods enhance the accuracy of CKM, especially in cases where channel measurements are spatially correlated but unevenly distributed.

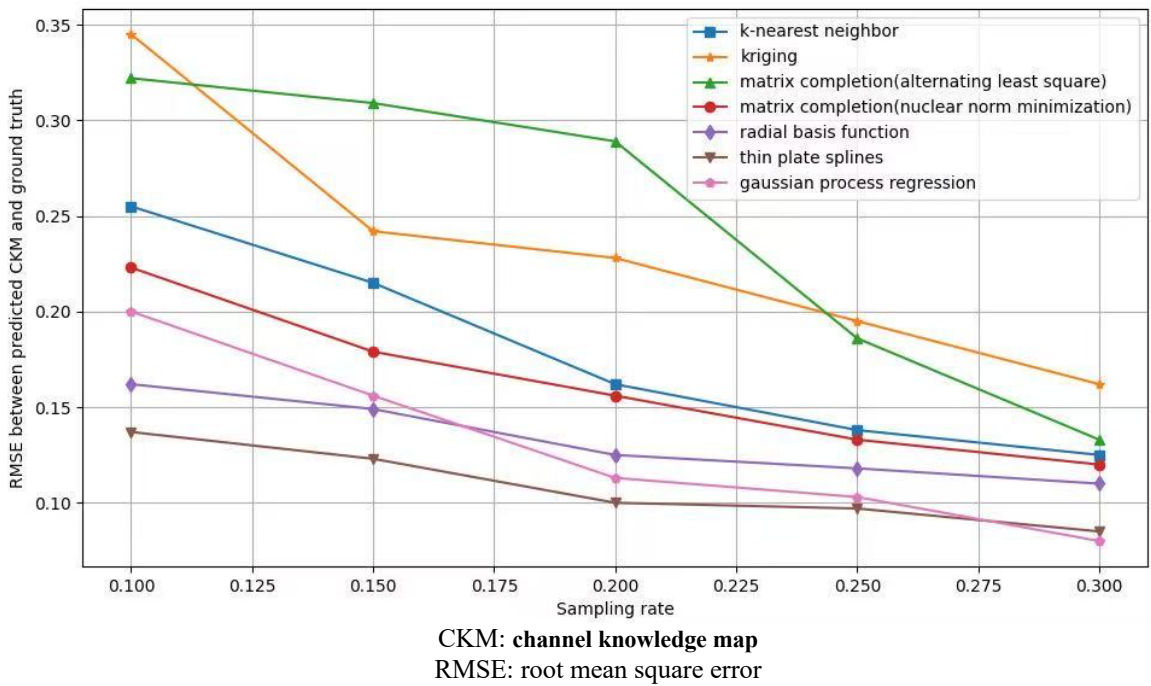

CKM: **channel knowledge map**
RMSE: root mean square error

Figure 3. Comparison of measurement-driven CKM construction methods

Fig. 3 compares the performance of several measurement-driven methods for constructing CKMs, using the received power map shown in Fig. 2b as an example, which is generated by Wireless Insite. The comparison is based on the root mean square error

(RMSE) between the predicted CKM and the ground truth values. The methods evaluated include the *k*-nearest neighbor, Kriging, matrix completion using alternating least squares and nuclear norm minimization, radial basis function, thin plate splines, and Gaussian process regression. The *x*-axis represents the sampling rate, which influences the amount of data used for interpolation, while the *y*-axis shows the RMSE, indicating prediction accuracy.

KNN and Kriging are intuitive and widely used but may suffer from limited accuracy when measurement data are sparse or highly variable. Matrix completion methods are effective in handling missing data but rely on low-rank assumptions that may not always hold. RBF and thin plate splines offer smooth interpolation, with the latter showing consistently low RMSE across sampling rates in Fig. 3. Gaussian process regression provides accurate predictions with uncertainty estimates but is computationally more demanding. The choice of interpolation and regression methods should consider both accuracy requirements and computational cost in practical CKM applications.

It is important to note that these results are based on a specific example case, and the performance of these methods may vary depending on various factors, such as the type of CKM, the environmental conditions (e.g., the distribution of buildings), and the specific locations of transmitters and receivers. Therefore, the selection of the appropriate method should be carefully considered based on the particular characteristics of the environment and network setup in practice.

**3.2 Model-Based CKM Construction**
Model-based CKM construction utilizes established propagation models to derive channel characteristics from environmental and systematic parameters. Unlike measurement-driven approaches that require extensive measurements, model-based methods use theoretical formulations to predict channel conditions at different locations. These methods can be broadly categorized into statistical channel models and deterministic channel models, with the latter primarily relying on ray tracing techniques.

1) Statistical model
Statistical channel models describe the wireless channel using probabilistic distributions derived from empirical observations and theoretical analysis. These models capture large-scale and small-scale fading effects, allowing CKM construction based on generalizable statistical properties rather than site-specific measurements.

For large-scale channel variations, path loss models, such as the COST-231 Hata[14] and Okumura models[27], estimate the average signal attenuation as a function of distance, frequency, and environmental factors. These models provide a coarse representation of CKM, making them suitable for initial coverage predictions in urban and suburban environments. Shadowing models, e.g., the log-normal shadowing model[28], account for signal fluctuations due to obstructions, incorporating randomness into path loss predictions.

More advanced models consider the distribution of environmental factors, such as building density, height, and urban layout, to refine path loss predictions. The Walfisch-Bertoni model[29] incorporates diffraction and reflection effects in dense urban environments, adjusting signal attenuation parameters based on the presence of obstacles. Similarly, recent studies have introduced geometry-based stochastic models, which approximate line-of-sight (LoS) and non-line-of-sight (NLoS) probability in urban environments by considering the statistical distribution of buildings and their impact on signal propagation[30]. These models reveal that the density and spatial distribution of buildings significantly impact signal behavior. Another approach extends stochastic probability models to air-to-ground (A2G) communications, analyzing LoS, ground specular, and building-scattering paths based on urban topology[31]. Furthermore, sub-THz statistical models have been developed to study spatial channel characteristics in urban microcells, focusing on spatial clustering and power distributions at high frequencies[32].

For small-scale fading, Rayleigh and Rician fading models describe the rapid variations in received signal strength caused by multipath propagation[33–34]. Rayleigh fading is commonly used in rich-scattering environments with no dominant LoS component, while Rician fading accounts for the presence of a strong LoS path. These models enable CKM construction that reflects the statistical behavior of fading effects, which is crucial for evaluating signal reliability in dynamic wireless environments.

While statistical models provide a computationally efficient means of CKM construction, they lack site-specific accuracy, as they rely on general assumptions rather than precise environmental information. As a result, they are often complemented by deterministic models that incorporate real-world physical conditions.

2) Deterministic model
Deterministic models predict wireless channel characteristics based on the physics of electromagnetic wave propagation, making them inherently environment-aware. These models incorporate detailed environmental features such as building layouts, vegetation, and material properties to accurately capture wave interactions, including reflection, diffraction, and scattering. Prominent deterministic approaches include ray tracing[15] and finite-difference time-domain (FDTD) simulations[35], both of which approximate solutions to Maxwell's equations. Ray tracing is commonly used due to its high-frequency approximation capabilities, whereas FDTD provides more precise but computationally intensive solutions based on discretized Maxwell's equations[36].

The distribution of buildings within a given environment significantly influences the accuracy of these deterministic models. For instance, ray tracing simulates the paths of electromagnetic waves by modeling interactions with environmental features. The placement, density, and geometry of buildings directly affect the reflection, diffraction, and scattering events considered in ray tracing simulations. Variations in building distribution can lead to significant differences in predicted signal paths, affecting the reliability of communication systems.

By leveraging precise environmental and electromagnetic characteristics, deterministic models offer highly accurate path loss predictions[37–39]. Recent advancement, such as improved geometric and material characterization techniques[40], has further enhanced model accuracy by reducing discrepancies between simulations and real-world measurements. However, a key challenge of deterministic models is their computational complexity, which makes large-scale, real-time applications difficult.

Several ray-based simulation methods have been developed to improve efficiency, including shooting and bouncing ray (SBR)[41] and vertical-plane-launch (VPL)[42]. Additionally, advanced techniques like intelligent ray tracing (IRT)[43] provide further acceleration, making deterministic modeling more feasible for CKM construction. Despite their computational demands, these models remain crucial for site-specific, high-accuracy channel prediction in next-generation wireless networks.

## 3.3 Measurement-Model Hybrid CKM Construction

Hybrid CKM construction methods integrate both measurement-driven techniques and theoretical propagation models to improve accuracy and efficiency. These approaches leverage the strengths of both paradigms: measurement-driven methods utilize machine learning to extract patterns from measurements, while model-based methods incorporate physical constraints to ensure consistency with wireless propagation principles. Recent research in this field has focused on two primary strategies for hybrid CKM construction.

1)  Computer vision approach

One major trend in hybrid CKM construction is leveraging computer-vision-based deep neural networks to process environmental and sparse measurement data as multi-channel inputs, treating CKM construction as an image-to-image translation task. This approach enables the extraction of spatial features from diverse data sources, including transmitter-receiver locations, building distributions, and limited channel measurements.

Computer-vision-based approach integrates 3D building maps, environmental features, and sparse channel measurements as multi-channel inputs into neural networks, allowing models to infer missing channel information and construct accurate CKMs. A widely used framework is RadioUNet[44], which extends the U-Net architecture by incorporating measurement-assisted inputs alongside environmental maps, improving prediction accuracy. Similarly, models such as EME-Net[45] and ACT-GAN[46] refine CKM predictions by leveraging deep neural networks trained on transmitter positions and building layouts. These architectures enhance generalization by learning structural patterns in radio maps while adapting to different urban environments. Subregional learning techniques[47] offer additional benefits by segmenting the channel gain map into subregions and applying specialized models to each, which is particularly effective in complex environments where traditional models struggle. These techniques allow for more accurate predictions by addressing regional propagation characteristics more effectively.

GAN-based models, such as SS-GAN[48] and RME-GAN[49], have also demonstrated significant improvement in CKM construction by generating realistic channel maps from incomplete data. For instance, the two-stage framework[50] first uses a radio map prediction GAN (RMP-GAN) to generate coarse radio maps based on environmental data, which are then refined with sparse measurement data through a correction GAN (RMC-GAN). This approach is highly relevant for CKM construction as it corrects predictions based on real-time measurements, reducing inaccuracies typically found in traditional models. These adversarial learning techniques effectively enhance CKM accuracy, even in scenarios where building information is incomplete, or transmitter locations are unknown.

Additionally, hybrid architectures incorporating variance prediction and uncertainty modeling have gained traction. A notable example is the dual-UNet framework[51], where two separate but identical U-Net models are trained in parallel: one to estimate received signal strength (RSS) values and another to predict variance maps that quantify uncertainty. This design allows CKM construction to incorporate confidence levels in its predictions, making it more robust to missing or inaccurate input data. Studies on Gaussian-based modeling[52–53] further highlight the importance of integrating statistical uncertainty into deep learning-based CKMs, particularly for urban and indoor wireless environments.

Overall, computer vision-driven CKM construction methods shown in Table 1 provide flexible, data-efficient alternatives to conventional interpolation and model-based techniques. By integrating spatial, spectral, and temporal information, these architectures enable highly accurate, scalable, and real-time CKM generation, paving the way for intelligent wireless network optimization.

Table 1. Overview of computer vision-based CKM construction models

| Model Name | CKM Type | Key Approach | Main Features |
|---|---|---|---|
| RadioUNet[44] | Channel gain map | U-Net (CNN-based) | ● Uses city map and transmitter location as input to estimate radio maps<br>● Incorporates physical simulation data for training<br>● Utilizes transfer learning to adapt simulated data to real-life scenarios |
| EME-Net[45] | Indoor RF-EMF exposure Map | U-Net (CNN-based) | ● Four input channels (red, green, blue, alpha) representing received power intensity<br>● Trained on Wi-Fi access points in realistic indoor environments |
| ACT-GAN[46] | Channel gain map | GAN (with AOT block, CBA module and T-Conv block) | ● Three input channels representing 3D building maps, transmitter locations, and environmental features<br>● Trained on sparse channel measurements and environmental data from urban areas<br>● Robust performance in scenarios with sparse discrete observations and unknown emission sources |
| Subregional learning-based CGM[47] | Channel gain map | MCNN-1D | ● Divides the map into subregions using a measurement-driven clustering approach<br>● Input channels: spatial coordinates of BS and sample points, and channel gain<br>● Trained on simulated channel data from a target area |
| SS-GAN[48] | RF coverage and interference map | GAN | ● Two input channels: RF coverage data and geographic data (elevation and building height)<br>● Trained on 4G LTE real-world data<br>● Uses sparsely self-supervised learning for weak supervision |
| RME-GAN[49] | Channel gain map | cGAN | ● Two-phase framework: Phase 1 integrates radio propagation models, Phase 2 captures local shadowing effects<br>● Trained on sparse RF measurements from 700 radio maps, including data from various urban regions like Ankara, Berlin, and Tel Aviv<br>● Inputs: sparse observations, transmitter locations, and urban maps |
| FPTC-GANs[50] | Channel gain map | RMP-GAN, RMC-GAN | ● Training data comes from real-world measurements and environmental information (e.g., transmitter positions, obstacle heights, etc.)<br>● Inputs: transmitter positions, obstacle top views, and empirical radio map<br>● First-predict-then-correct approach (RMP-GAN for initial prediction, RMC-GAN for correction) |
| GAN-CRME[51] | Channel gain map | cGAN | ● Inputs: distributed RSS samples and geographical map<br>● Trained on a dataset with RSS samples and geographical map information |
| SSSP[52] | Channel gain map | U-Net | ● Inputs: signal strength measurements, 3D map of the environment (urban)<br>● Training data: signal strength data generated using wireless InSite ray-tracing software from simulated environments (45 urban environments)<br>● The model does not require transmitter location or statistical channel models |
| REM-U-Net[53] | Channel gain map | U-Net | ● Inputs: building height maps, building layout maps, and LoS maps |

| | | | • Trained on the RadioMap3DSeer dataset with simulated data from 701 city maps<br>• Uses LoS maps as additional input to improve prediction accuracy |
|---|---|---|---|

AOT: aggregated contextual transformation
CBA: convolutional block attention
cGAN: conditional generative adversarial network
CGM: channel gain map
CKM: channel knowledge map
CNN: convolutional neural network
CRME: cooperative radio map estimation
EMF: electromagnetic field
FPTC: first-predict-then-correct
GAN: generative adversarial network
LoS: line-of-sight
MCNN: modular convolutional neural network
RMC: measurement data correction
RME: radio map estimation
RMP: radio map prediction
SSSP: spatial signal strength prediction

2) Calibrated ray tracing

Calibration using real-world measurements has become a key research focus to improve ray tracing performance, since the precise environmental parameters such as material permittivity, reflection coefficients, and scattering effects are difficult to obtain. Several techniques have been developed to refine material properties, incorporate diffuse scattering effects, and adjust propagation parameters based on empirical data.

One common calibration strategy is tuning material properties using real-world measurements. In Ref. [54], the relative permittivity of materials in urban microcell (UMi) environments at 28 GHz was fine-tuned, reducing errors in path loss estimation. Similarly, a linear interpolation approach was used in Ref. [55] to estimate the dielectric constant of concrete at 28 GHz based on measured values at 5.2 GHz and 60 GHz, ensuring more accurate reflection loss modeling. Additionally, empirical ray tracing models have been validated against 73 GHz street canyon measurements, refining reflection loss calculations based on the incident angle[56].

Another key challenge in ray tracing is capturing diffuse scattering effects, which become significant at mmWave and THz frequencies. Calibration against 28 GHz urban directional channel measurements in Ref. [57] involved adjusting scattering coefficients and incorporating an angular spread correction factor, improving received power predictions. At THz frequencies, extensive indoor ultra-broadband measurements were conducted in Ref. [58], leading to the development of a frequency-dependent scattering model that reduced errors in delay spread estimation. In an office setting at 60 GHz, multipath component gains were analyzed and adjusted to improve model reliability[59]. The NYURay ray-tracing calibration method simplifies the process by assuming angle-independent reflection, enabling a closed-form least squares optimization to align simulated multipath power with real-world measurements[60]. Instead of iterative tuning, the method directly optimizes reflection and penetration losses in a logarithmic scale, improving efficiency while maintaining accuracy.

In recent years, research has investigated the integration of neural networks with traditional ray tracing frameworks. In Ref. [61], the authors apply neural networks to the interaction calculation module, utilizing neural networks to predict the output direction and losses of each interaction. The architecture consists of two main components: the Spatial Network and the Material Network. The Spatial Network processes the spatial characteristics of the ray's path, while the Material Network accounts for the material properties influencing the ray's behavior. However, this framework does not consider high frequency in future communication applications, which limits its effectiveness within certain application ranges. Apart from this, Ref. [62] proposes a learnable wireless digital twin, which, similar to ray tracing frameworks, integrates neural networks. This framework uses a single entity representing each object within the environment, constructing a neural network to encode its electromagnetic property one by one, which results in an improved accuracy for channel modeling. However, the design of neural networks in large-scale systems also increases complexity and the requirement for computational resources.

3.4 Summary of Advantages and Disadvantages of CKM Construction Methods

The various CKM construction methods each have their unique advantages and limitations. Measurement-based methods, such as KNN and Kriging, are highly effective when sufficient real-world measurement data are available, providing high-level accuracy. However, their performance heavily relies on the density and quality of the data. In cases where data are sparse or irregular, their effectiveness can be reduced. Model-based methods, including statistical models and ray tracing, are useful when measurement data are limited, as they rely on theoretical models. While they can be efficient, their accuracy may be lower in complex or rapidly changing environments, and may fail to capture fine-grained variations.

Hybrid methods that combine measurement-driven approaches with theoretical models strike a balance between accuracy and efficiency. These methods can offer improved performance in dynamic environments by leveraging both empirical data and physical models, but they are computationally more demanding. The choice of method largely depends on the available data, computational resources, and the complexity of the environment, with measurement-based methods often being preferred for high-accuracy scenarios and model-based methods being more suited for situations with limited measurements.

## 4 Applications of CKM

### 4.1 Integrated sensing and communication

CKM plays a critical role in integrated sensing and communication (ISAC) systems, where it bridges the gap between sensing and communication systems. By providing detailed, location-specific channel information, CKM not only enhances localization but also improves communication performance in dynamic environments. Unlike traditional methods that rely solely on measurements like RSS, CKM offers additional features such as time of arrival and AoA, which are crucial for both localization and beamforming optimization[63–64]. In terms of communication, CKM helps in adjusting the communication links between base stations and mobile users by integrating sensing data, such as dynamic environmental changes, into the channel model[65]. This integration allows for more efficient resource allocation, interference management, and adaptive beamforming. For example, in unmanned aerial vehicle (UAV) systems, CKM can simultaneously support both the localization of UAVs and the optimization of their communication links with ground stations by using real-time channel state information[66]. Additionally, CKM allows for dynamic sensing of moving objects, such as vehicles and pedestrians, and enables real-time updates of the communication network based on the sensed data, optimizing the overall system performance[67].

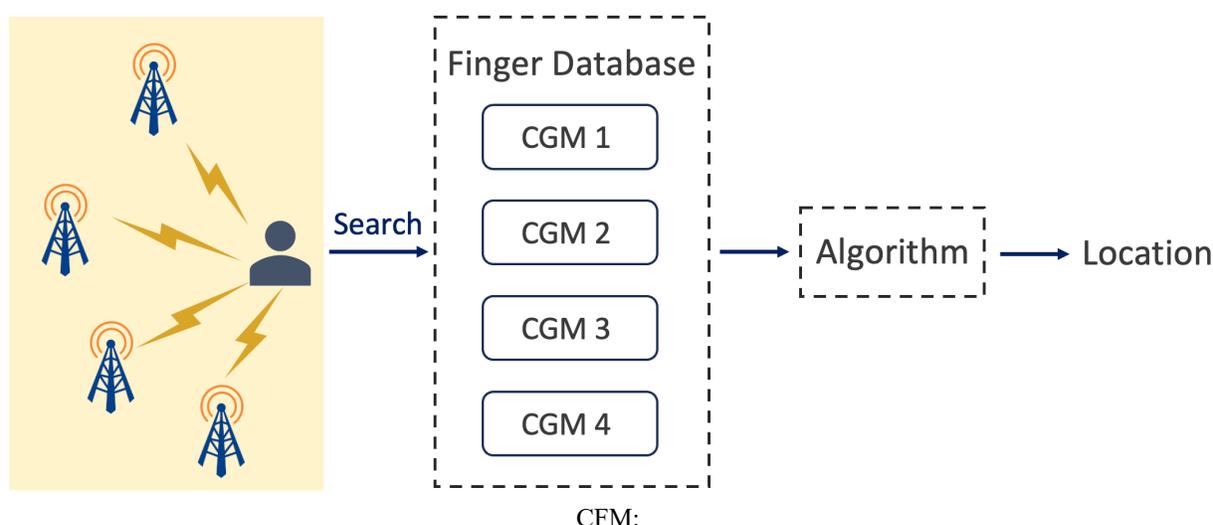

CFM:
Figure 4. Fingerprint localization using CGM in integrated sensing and communication systems

Fig.4 illustrates the application of CGMs in ISAC systems, specifically within the fingerprint localization algorithm. In this approach, the user's signal characteristics are collected and matched against multiple CGMs stored in a Finger Database. The algorithm processes this data to estimate the user's location based on the similarity of the measured signal to the stored CGMs. This example highlights how CKMs, including various types beyond CGMs, can enhance both the sensing and communication capabilities in ISAC systems, offering a more accurate and efficient means for location tracking in complex environments.

### 4.2 UAV Trajectory Optimization

CKM plays a critical role in enhancing UAV trajectory optimization in 6G networks, both for cellular-connected UAVs and UAV-assisted communication systems. By utilizing rich, location-specific channel data, CKM helps construct real-time SINR maps that account for both channel gain and interference. In the cellular-connected UAV scenario, CKM aids in constructing SINR maps based on channel gain and interference, allowing for the design of flight paths that minimize outage probability and mission latency while ensuring continuous communication with ground base stations (BSs). Unlike traditional methods based on deterministic LoS or stochastic channel models, which fail to account for LoS blockages, CKM enables the UAV to navigate areas with high interference or blockages while maintaining a reliable signal-to-interference plus noise ratio (SINR) throughout the path[68]. In UAV-assisted systems, where a UAV serves multiple users, CKM helps optimize the trajectory by identifying paths with strong

A2G channels, thus enhancing overall system performance. This method ensures better coverage for all users, compared to conventional designs that focus on a single user's location. Additionally, combining reinforcement learning with CKM allows the UAV to dynamically adjust its trajectory based on real-time user distribution and channel quality data, improving communication efficiency across the system[69–70].

### 4.3 Hybrid Beamforming

CKM significantly enhances hybrid beamforming for mmWave massive multiple-input multiple-output (MIMO) systems, offering a more efficient approach to reduce the complexity and overhead typically associated with traditional training-based channel estimation[71–72]. In these systems, hybrid analog-digital beamforming uses a combination of analog and digital beamforming techniques at both the transmitter and receiver to manage multiple data streams effectively. Traditional beamforming approaches often rely on extensive channel state information and require significant training to estimate the full MIMO channel matrix, which becomes more challenging as the number of antennas increases. With the integration of CKM, such as the channel angle map (CAM) and beamforming indicator map (BIM), the need for extensive training is minimized, as the system can use location-specific channel information, including AoA, , and path loss, that is directly derived from the environment. This enables training-free or light-training beamforming, where the accuracy of these designs depends on the precision of user location and environmental factors like scatterer movement. While limited training may still be beneficial to refine the system, CKM improves beamforming efficiency, particularly in dynamic environments with multiple users or interference, optimizing data throughput and SINR over large antenna arrays.

### 4.4 Base Station Placement

In base station deployment, CKM applications assist in optimizing base station placement strategies to enhance network coverage and performance. By constructing detailed CKMs, one can accurately assess signal strength, interference levels, and coverage areas at various locations, thereby determining optimal base station positions and configurations. For instance, in low-attitude environments, methods for deploying multiple aerial base stations utilize binary CKMs, to optimize base station layouts to meet the communication needs of different areas[73]. Additionally, principles for deploying ultra-wideband (UWB) indoor positioning system base stations emphasize the importance of CKMs. By analyzing indoor channel characteristics, reasonable base station placement can improve positioning accuracy and system performance[74]. In summary, CKM applications in base station deployment support more intelligent base station placement decisions by providing precise channel information, thereby enhancing network coverage and service quality.

### 4.5 Resource Allocation

In 6G networks, particularly for ultra-reliable low-latency communication (URLLC) in mission-critical IoT systems, CKMs can be used to optimize resource allocation by adapting transmission control policies. These policies aim to meet the stringent quality of service (QoS) requirements of URLLC while minimizing the transmit power. By utilizing CKMs, which provide channel gain statistics for various locations within a target area, transmission control can be optimized without the need for real-time, costly channel state information. A notable approach including power scaling based on CKM data was proposed[75], where location-specific transmission parameters are adjusted to maintain a target delay violation probability delay violation probability across all devices. This method ensures that devices in varying conditions can still operate within the desired reliability and latency constraints. Additionally, meta-reinforcement learning techniques have been employed to further enhance adaptability, enabling rapid policy adjustment across different environments with minimal retraining. This combination of CKM-driven power scaling and meta-learning offers a scalable solution to resource allocation in URLLC systems.

## 5 Open Problems and Future Directions

### 5.1 Localization Accuracy and Robustness of CKM

One major challenge for CKM-based systems is the dependence on high-precision localization data. Since CKMs rely on accurate location information to construct location-specific channel knowledge, errors in localization can directly affect the performance and robustness of the system. Inaccurate positioning data, such as from GPS or environmental obstructions, can distort the generated channel map, leading to suboptimal outcomes in applications like localization, beamforming, and interference management. To address this issue, future research must focus on maintaining CKM robustness in the presence of localization inaccuracies. This may involve using machine learning techniques to compensate for errors or applying sensor fusion methods to combine various positioning sources. Additionally, techniques such as spatial smoothing or interpolation can help mitigate the impact of small localization errors, ensuring CKM construction remains reliable even with less precise location data.

### 5.2 Incorporating Material Properties for Enhanced CKM Accuracy

Another significant challenge for CKM-based systems is efficiently incorporating detailed environmental information, particularly the impact of various material properties on CKM. While most existing methods primarily rely on geometric information, such as building shapes or the layout of obstacles, they often overlook how the materials of these objects (e.g., walls, windows, or furniture) affect the propagation of radio waves. Different materials, with varying electromagnetic properties, can significantly influence path loss, reflection, and scattering, which in turn affect the accuracy of the CKM. To address this, future research should focus on integrating material-specific data with CKM construction. This may involve leveraging detailed environmental sensing, such as materials' electromagnetic characteristics, or using machine learning to predict the impact of materials on the channel. Combining geometric and material information will improve the fidelity of CKMs, making them more reflective of real-world conditions and enhancing applications such as beamforming and localization in complex environments.

### 5.3 Improving Generalization with Efficient Neural Network Architectures for CKM Construction

A third key challenge is enhancing the generalization ability of CKM construction, particularly when using neural network-based methods. Currently, most neural network models require training on a large variety of scenarios to achieve robust performance. However, this process can be time-consuming and computationally expensive. The ability to design more efficient neural network architectures that can be trained on fewer scenarios while maintaining strong performance across a wide range of environments is crucial. To tackle this, future research should focus on developing models that require minimal training data, perhaps by using transfer learning, domain adaptation, or few-shot learning techniques. These approaches may enable neural networks to generalize better and perform well across different deployment scenarios, making CKM-based systems more scalable and effective for real-world applications, even with limited training data.

### 5.4 Continuous CKM Updates with Real-Time Data

A crucial challenge for CKM-based systems is how to continuously update the CKM with new data, ensuring its accuracy and relevance over time. In dynamic environments, the wireless channel is constantly changing due to factors like mobility, environmental alterations, and user behavior. To maintain an up-to-date CKM, it is essential to integrate new measurements and real-time data effectively. This could be achieved through techniques such as incremental learning and online learning, where the CKM model is continuously updated as new data are acquired, without the need to retrain from scratch. Additionally, sensor fusion methods can be employed to combine data from different sources, such as measurement devices, UAVs, and sensors for Internet of things, providing a more comprehensive and accurate representation of the environment. By incorporating these approaches, CKM systems could adapt in real-time to changing conditions, ensuring that they remain accurate and reliable for various applications.

### 5.5 CKM in 6G for Robotics

In the future, the application of CKMs in robotics will play a crucial role in advancing 6G technologies. As robotics continues to evolve, key aspects such as accurate channel modeling, enhanced localization accuracy, and efficient sensing capabilities will become increasingly important. CKMs can significantly contribute to these areas by providing detailed, environment-aware channel information, enabling robots to navigate and interact more effectively in dynamic environments [76]. Through the integration of CKMs with 6G networks, robots can benefit from more reliable localization and real-time sensing, improving their ability to adapt to changing conditions and interact with both humans and other devices seamlessly. Efficient use of the communication channel will also be vital for optimizing robot performance, ensuring low latency and high throughput for tasks such as autonomous control, monitoring, and remote operation.

## 6 Conclusions

This paper has provided an in-depth overview of CKMs and their transformative role in 6G wireless networks. CKMs represent a paradigm shift from environment-agnostic communication to environment-aware communication, allowing for more efficient channel estimation and resource allocation. Through various CKM construction methods from measurement-based and model-based techniques to hybrid approaches, researchers have demonstrated the potential to improve channel knowledge accuracy, particularly in complex environments. The applications of CKMs, including ISAC systems, beamforming, UAV trajectory optimization, base station placement and resource allocation, highlight their broad impact on network performance and optimization.

As 6G technologies evolve, the integration of CKMs with advanced systems such as reconfigurable intelligent surfaces (RIS), mmWave communications, and machine learning-based adaptive resource management holds great promise. Combining CKMs with these technologies can dynamically optimize the communication environment, improve coverage in challenging areas, and enable real-time adaptation to network changes, further enhancing system efficiency and reliability. However, despite their promising applications, several challenges remain in the development and deployment of CKM systems. Future research should focus on improving CKM robustness in scenarios with imprecise localization data, better integrating material-specific environmental information and enhancing the generalization capabilities of neural network models. Moreover, continuously updating CKMs with real-time data will be crucial for maintaining their relevance and accuracy in dynamic environments.

Acknowledgement:

This work was supported in part by the National Natural Science Foundation of China under Grants Nos. 62431014 and 62271310, and in part by the Fundamental Research Funds for the Central Universities of China.